# 3D PET image reconstruction based on Maximum Likelihood Estimation Method (MLEM) algorithm

Artur Słomski[1,#], Zbigniew Rudy[1,#], Tomasz Bednarski[1], Piotr Białas[1], Eryk Czerwiński[1], Łukasz Kapłon[1,2], Andrzej Kochanowski[2], Grzegorz Korcyl[1], Jakub Kowal[1], Paweł Kowalski[3], Tomasz Kozik[1], Wojciech Krzemień[1], Marcin Molenda[2], Paweł Moskal[1], Szymon Niedźwiecki[1], Marek Pałka[1], Monika Pawlik[1], Lech Raczyński[3], Piotr Salabura[1], Neha Gupta-Sharma[1], Michał Silarski[1], Jerzy Smyrski[1], Adam Strzelecki[1], Wojciech Wiślicki[3], Marcin Zieliński[1], Natalia Zoń[1]

[#]corresponding author, email: artur.slomski@gmail.com
[#]corresponding author, email: zbigniew.rudy@uj.edu.pl

[1]Faculty of Physics, Astronomy and Applied Computer Science, Jagiellonian University, 30-059 Kraków, Reymonta 4 Street, Poland

[2]Faculty of Chemistry, Jagiellonian University, 30-060 Kraków, Ingardena 3 Street, Poland

[3]Swierk Computing Centre, National Centre for Nuclear Research, 05-400 Otwock-Swierk, Soltana 7 Street, Poland

**Abstract:** Positron emission tomographs (PET) do not measure an image directly. Instead, they measure at the boundary of the field-of-view (FOV) of PET tomograph a sinogram that consists of measurements of the sums of all the counts along the lines connecting two detectors. As there is a multitude of detectors build-in typical PET tomograph structure, there are many possible detector pairs that pertain to the measurement. The problem is how to turn this measurement into an image (this is called imaging). Decisive improvement in PET image quality was reached with the introduction of iterative reconstruction techniques. This stage was reached already twenty years ago (with the advent of new powerful computing processors). However, three dimensional (3D) imaging remains still a challenge. The purpose of the image reconstruction algorithm is to process this imperfect count data for a large number (many millions) of lines-of-responce (LOR) and millions of detected photons to produce an image showing the distribution of the labeled molecules in space.

## Introduction

Positron emission tomography (PET) is a grown up technology used for medical imaging whose importance is still rapidly increasing. There is an established appreciation of the significance of the functional (as opposed to anatomical obtain e.g. via X-ray examination) information that is provided by PET, in particular of its value for the purposes of medical diagnosis and monitoring of the response to therapy. The essentials task in PET is to reconstruct a source distribution i.e. to obtain an accurate image of the radioactivity distribution throughout the patient. This is done in order to extract metabolic information about the patient body. PET imaging is unique in that it shows the chemical functioning of tissues "in vivo", while common imaging techniques – such as X-ray – show structure of tissues.

The means are as follow. One labels the chosen molecule (ligand) with a radioactive atom (i.e. one substitutes a radiotracer) and administers certain amount of the labeled molecules to the patient. The choice depends on the metabolic process of interest. The labeled molecules follow their specific biochemical tracts inside the patient body. The radioactive atoms (or rather their nuclei) used as labels are unstable β+ emitters and undergo radioactive decay at random directions, leading to the emission of positrons. A positron emitted during the radioactive decay process annihilates with an electron in tissue and as a result a pair of gamma quanta is emitted. The two gamma quanta fly back-to-back i.e. in opposite directions and can be recorded outside the patient body by scintillation detectors.

Every detected pair of quanta forms line-of-responce (LOR). Austrian mathematician Johann Radon [1,2,3] proved that from such projections if they are sufficiently numerous one can reconstruct radiation intensity (problem is well-posed). However, the solution does not have a closed-form expression. Numerical methods are required. Nowadays modern approach consists in iterative algorithms derived from Maximum Likelihood Estimation Method (MLEM).

The naïve reconstruction algorithm used to calculate the radioactivity distribution from the projections is based on counting activity. Algorithm adds activity for each pixel along an LOR detected by a detector pair. The process is repeated for all measured LORs, resulting in an image (discretized distribution of radiation intensity) of the original object. Such reconstructed image contains streak artefacts and is blurred.

### Two-dimensional imaging

Two-dimensional PET imaging considers only lines of response (LORs) lying within a specified imaging plane. The LORs are organized into sets of projections i.e. line integrals are calculated for all r for a fixed direction φ (see Figure 1). The collection of all projections as a two dimensional function of r and φ forms a sinogram in (r, φ) representation. The measured counts in the projection sinogram corresponding to the calculated r are added to the (x, y) pixel in the reconstruction matrix. This is repeated for all projection angles (Figure 1).

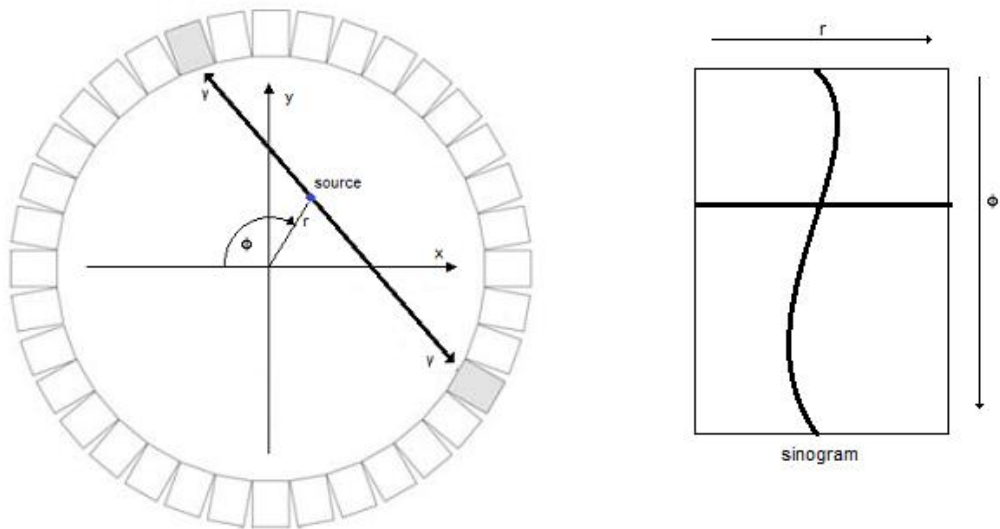

Figure 1. Schematic diagram of PET. Two gamma rays emitted as a result of positron annihilation are detected by two detectors.

The lines connecting the detectors (left side of the Figure) is described by coordinates (r, ϕ) and represented on sinogram (right side).

It is possible to reconstruct a whole 3D volumetric object by repeating the 2-D data acquisition for multiple axial (in z direction) slices, although procedure is tedious  When the sinogram for each value of z is reconstructed, one can stack the image planes together one after the other  to form a three-dimensional image. Although this can be considered as form of three-dimensional imaging, it is different from the three-dimensional acquisition model described in the next section. There is a handful of effective 2D iterative procedures for imaging [5].

## Three-dimensional imaging
Fully 3-D measurements require more storage of data. As a result, reconstruction becomes more computationally intensive. The solution is to use iterative methods such as MLEM (maximum likelihood expectation-maximization).

The diagram in Figure 2 shows the basic procedure for using an iterative algorithm.

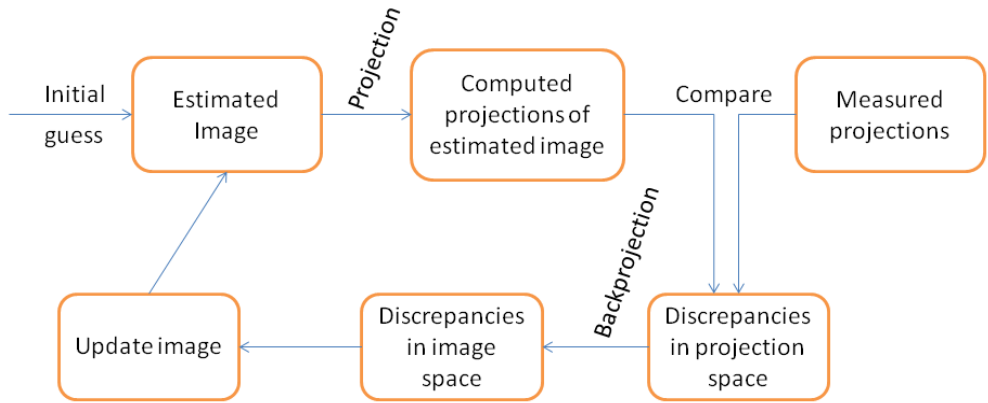

Figure 2. Flow chart of iterative image reconstruction method.

The initial estimate of the image in an iterative algorithm is usually a uniform distribution . The projections are computed from the image and compared with the measured projections. If there is a difference between the estimated and measured projections, corrections are made to improve the estimated image, and a new iteration is performed to assess the convergence between the estimated and measured projections. Iterations are continued until a reasonable agreement between the two sets of projections is achieved.

The MLEM reconstruction is given by [6]:

$$\lambda_j^{k+1} = \frac{\lambda_j^k}{\sum_i^m C_{ij}} \sum_i^m \frac{C_{ij}}{\sum_j^m C_{ij} \lambda_j^k}$$

where:

$\lambda_j^k$ – value of reconstructed image at the pixel j for the k-th iteration,

k – iteration number,

j – pixel number,

i – projection's bin number,

Cij – probability of detecting an emission from the pixel j in projection's bin i.

In three-dimensional PET imaging, one acquires all LOR also ones lying on 'oblique' imaging planes. Fully 3-D mode is used to increase sensitivity (by means of increasing of number of measured LORs) and thus to lower the statistical noise associated with photon counting improving the signal-to-noise ratio in the reconstructed image. In three-dimensional reconstruction one must expand the projection coordinates for another dimensions to transform line of response from (x, y, z) coordinates.

One of the ways to represent a projection for three-dimensional reconstruction (i.e. to label the bins in which measured LORs are counted) is to use projection's coordinates system (r, θ, φ, sign wekx, sign weky, sign wekz) where:

r – distance from the origin of the coordinate system,

θ - angle between LOR and positive half of axis OZ,

φ - angle between LOR projection onto XY plane and negative half of axis OX,

sign wekx – sign of a component x of distance vector r sign weky – sign of a component y of distance vector r sign wekz – sign of a component z of distance vector r.

Algorithm with that implemented projection's coordinate system is convergent as shown in Figure 3 below, which represents the comparison of images by SSIM method depending on the number of iterations (it should be mentioned that algorithm works well although the bins in projections space are not described uniquely).

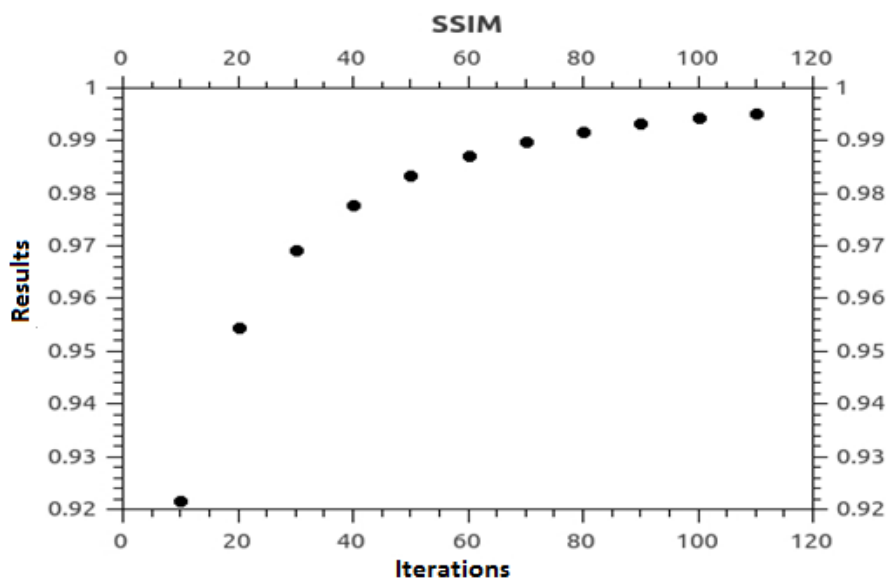

**Figure 3.** Algorithm convergence proof.
Horizontal i.e. abscissa axis: number of iterations. Vertical i.e. ordinate axis: value of structural similarity metric SSIM (SSIM measures spatial correlation between the pixels of the reference and test images to quantify the degradation of an image's structure, SSIM value is equal 1 only if two images are identical in considered pixels [4] ).

## Results

In this section examples of algorithm results are shown. First (Figure 4) original object i.e. the phantom is presented. It is assumed that the phantom radiates uniformly (Monte Carlo simulations were used in order to obtain 100 million of LORs). The phantom forms cylinder placed between two square bases. The bases are slightly larger than the cylinder ring (Figure 4). The cylinder is empty inside with central rod of rectangular section connecting the bases The cylinder is placed in three dimensional space with the bases parallel to xy plane. In the Figures 5,6,7 the results of the imaging algorithm are shown after 20, 50, 300 iterations, proving convergence. Really, the structure of the phantom is reproduced by imaging algorithm that is steered by simulated data. Cross-section at different levels of z coordinate for the reconstruction image are shown, for two-dimensional sections parallel to xy plane, a) for z equal 5 i.e. the section goes just through square basis of the cylinder, b) for z equal 6, the section shows the walls of the cylinder and rod in the center.

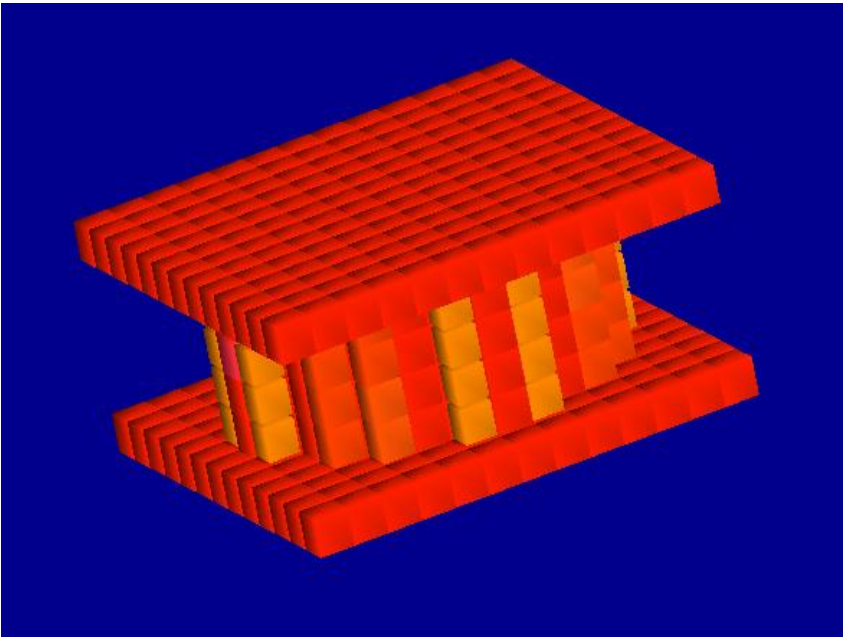

Figure 4. The shape of original image (phantom).

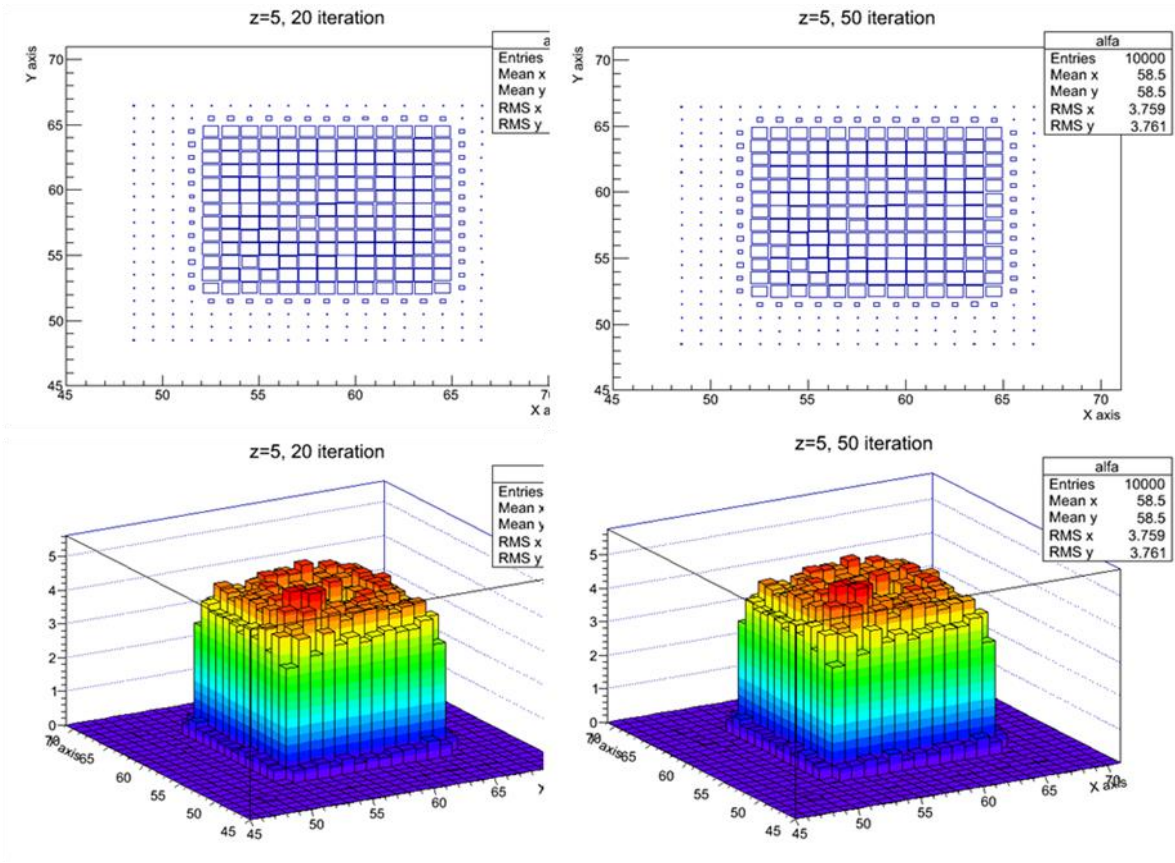

Figure 5. Reconstruction image shown for section at z=5 i.e. through square base of phantom.
Number of iterations: 20 and 50.

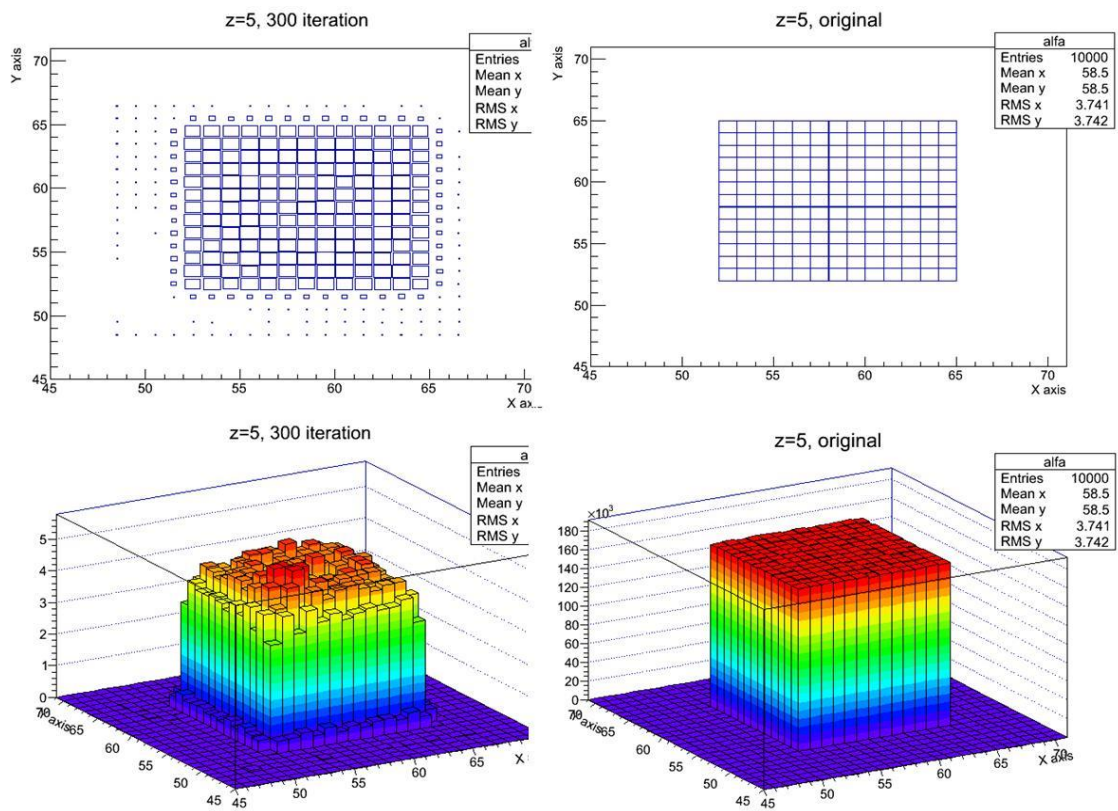

Figure 6. Reconstruction image shown for section at z=5 i.e. through square base of phantom. Upper left and lower left part of the Figure: result after 300 iterations. Upper right and lower right part of the Figure: one presents how the ideal image reconstruction should look like.

Figure 7. Reconstruction image shown for section at z=6 of phantom.

Number of iterations: 20, 50, 300. Walls and the central rod of phantom are reconstructed in a very good way. Additionally (plot entitled "original") it is presented how the ideal image reconstruction should look like at z=6.

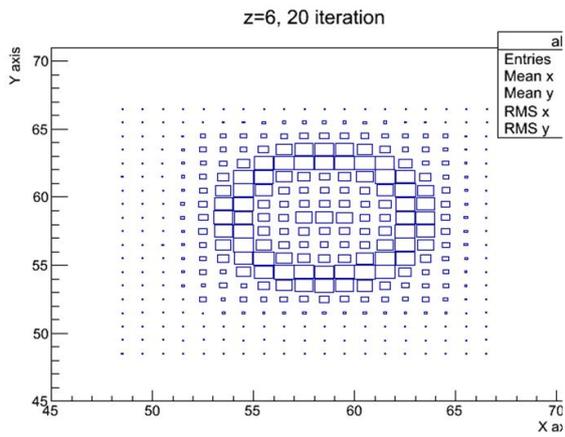
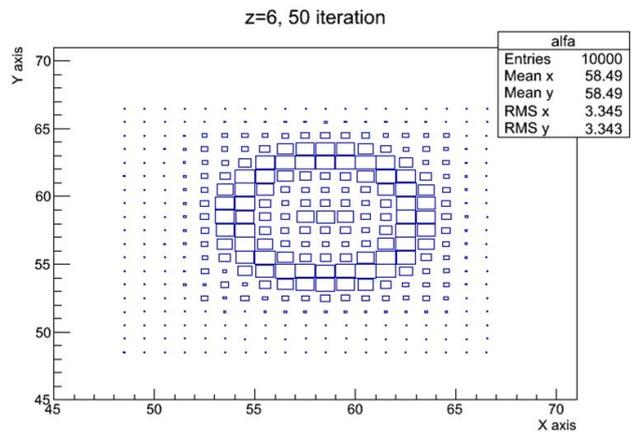
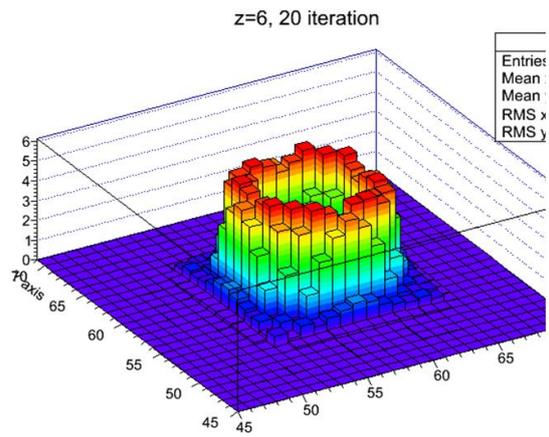
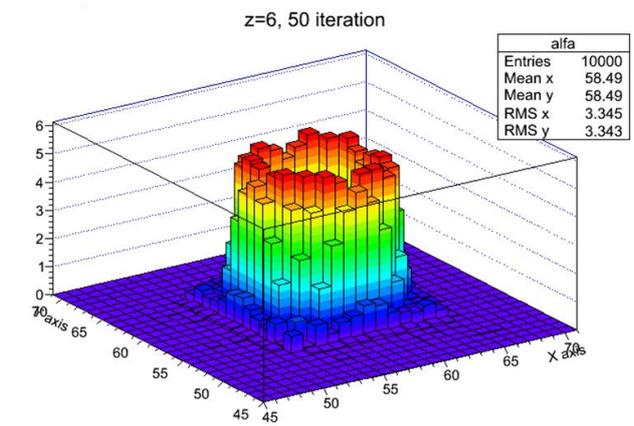
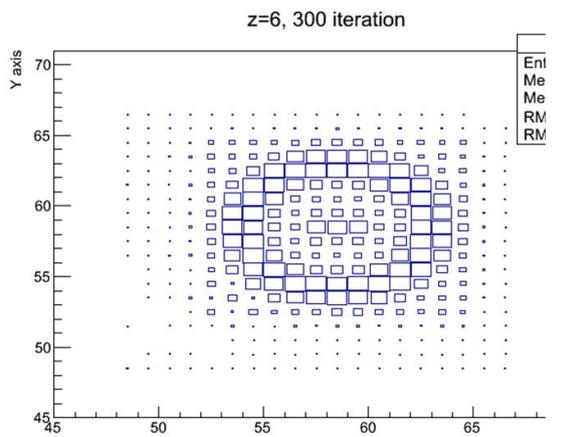
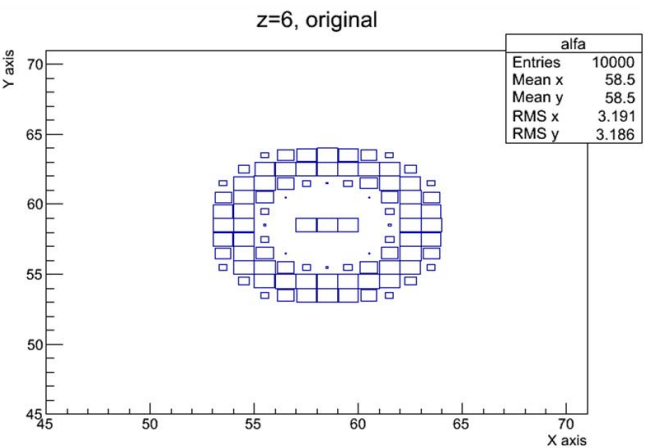
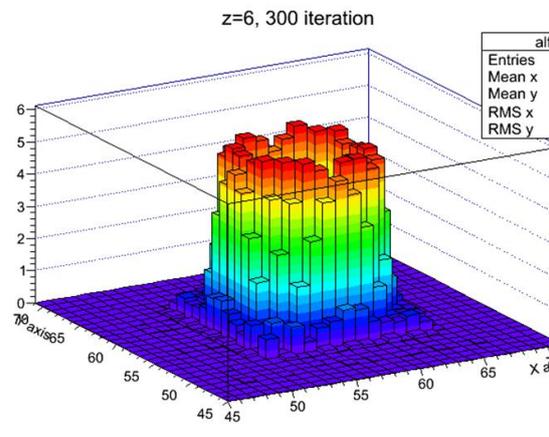
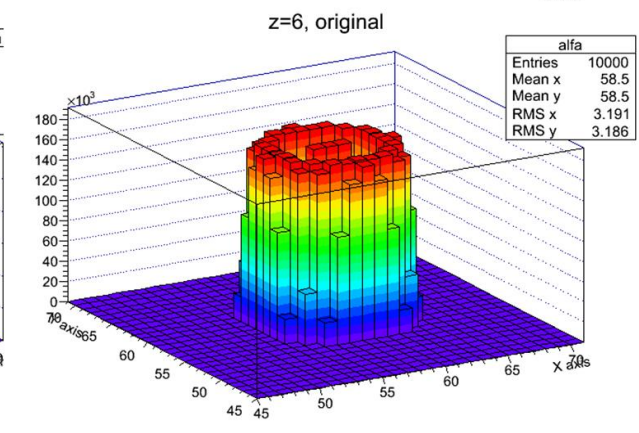

## Conclusions

The presented version of 3D image reconstruction algorithm, although it describes the bins of projection space only approximately (projection space stores information about accumulated LORs), works very well. It reaches quickly advanced degree of convergence.

As concerns statistical reconstruction methods, the success of presented algorithm enable us to state that statistical reconstruction methods seem to be a reasonable choice. Many assumptions about the noise can be made, but for the emission data the Poisson model (as concerns properties of distribution of emission) seems to be most adequate.

Nice feature of presented iterative (update) equations is that positivity constraint is automatically satisfied (reconstructed image, its pixels in radiation space should not have intensity smaller than 0). It should be mentioned that presented algorithm is quite immune for case when part of the gamma detectors for any reason is off; only LOR that were measured influence the result of the algorithm; signal-to-ratio may suffer, but no artefacts are formed.

It is often claimed that expectation maximization methods (EM) have drawback: noisy images are obtained from over-iterated reconstructions (over-iterated reconstruction may happen if unwieldy stopping rule is used). Our version of expectation maximization based algorithm is free from such unwanted behaviour.

## Acknowledgments

We acknowledge technical and administrative support by M. Adamczyk, T. Gucwa-Rys, A. Heczko, M. Kajetanowicz, G. Konopka-Cupiał, J. Majewski, W. Migdał, A. Misiak, and the financial support by the Polish National Center for Development and Research through grant INNOTECH-K1/IN1/64/159174/NCBR/12, the Foundation for Polish Science through MPD programme and the EU and MSHE Grant No. POIG.02.03.00-161 00-013/09.

## References

1. Radon J, Ueber die Bestimmung von Funktionen durch ihre Integralwerte laengsbestimmter Mannigfaltigkeiten, Ber. Verb. Sächs. Akad.Wissenschaften Lepzig, Math.-Nat. Kl. 1917; 69:262-77.

2. Herman G T, Image Reconstruction from Projections, Academic Press New York, 1980.

3. Smith K T, Keinert F, Mathematical foundations of computed tomography, Applied Optics 1985; 24:3950-57.

4. Wang Z, A. Bovik C, Sheikh H R, Simoncelli E P, Image quality assessment: from error visibility to structural similarity, IEEE Trans. Image Process. 2004; 16:600-12.

5. Parra L, Barrett H, List mode likelihood: EM algorithm and image quality estimation demonstrated on 2D PET, IEEE Trans. Med. Imag. 1998; 17:228-35.

6. Yokoi T, Shinohara H, Hashimoto T, Yamamoto T, Niio Y, Proceedings of the Second International Workshop on EGS, 8.-12. August 2000, Tsukuba, Japan, KEK Proceedings 2000; 20:224-34.